\documentclass{article}

\usepackage[utf8]{inputenc}
\usepackage{amssymb,amsmath, amsfonts,eurosym,geometry,ulem,graphicx,caption,color,setspace,sectsty,comment,footmisc,caption,pdflscape,subfigure,array,hyperref, csquotes, tabularx, fullpage, rotating, multirow, lscape, fancyhdr, booktabs, indentfirst}
\usepackage[linesnumbered,ruled,vlined]{algorithm2e} 
\usepackage[longnamesfirst]{natbib}
\usepackage{float}
\usepackage{caption}
\begin{document}

\begin{titlepage}
\title{\noindent{Beating the House:} \\ Identifying Inefficiencies in Sports Betting Markets}
\author{Sathya Ramesh\thanks{We would like to thank David Scott, Kathy Ensor, Michael Gordy, Benjamin Gardner, and seminar participants from Rice University and the University of Connecticut for their suggestions and encouragement. Additionally, we thank the Center for Computational Finance and Economic Systems at Rice University for research funding and support. The opinions expressed here are our own, and do not reflect the views of the Board of Governors or its staff. Send correspondence to Sathya Ramesh, Risk Analysis Section, Board of Governors of the Federal Reserve System, 20th \& C Streets, NW, Washington, DC 20551, Telephone: (202) 973-6124, e-mail: sathya.p.ramesh@frb.gov} \\ \small{Federal Reserve Board} \\ \small{sathya.p.ramesh@frb.gov} \and Ragib Mostofa\footnotemark[1] \\ \small{Rice University} \\ \small{rm48@rice.edu} \and Marco Bornstein\footnotemark[1] \\ \small{University of Maryland} \\ \small{marcob@umd.edu} \and John Dobelman\footnotemark[1] \\ \small{Rice University} \\ \small{dobelman@rice.edu}}
\date{\today}
\maketitle
\begin{abstract}
Inefficient markets allow investors to consistently outperform the market. To demonstrate that inefficiencies exist in sports betting markets, we created a betting algorithm that generates above market returns for the NFL, NBA, NCAAF, NCAAB, and WNBA betting markets.  To formulate our betting strategy, we collected and examined a novel dataset of bets, and created a non-parametric win probability model to find positive expected value situations. As the United States Supreme Court has recently repealed the federal ban on sports betting, research on sports betting markets is increasingly relevant for the growing sports betting industry.\\
\vspace{0in}
\vspace{0in}\\

\bigskip
\end{abstract}
\setcounter{page}{0}
\thispagestyle{empty}
\end{titlepage}
\pagebreak \newpage

\doublespacing

\section{Introduction} \label{sec:introduction}

\setlength\parindent{24pt}

\indent Casino games are designed to make gamblers choose from a range of negative expected value outcomes. In the short run, a gambler may profit; however, in the long run, the house always wins due to the Law of Large Numbers. The sentiment of the house inevitably winning is paralleled in financial markets by the strong form of the Efficient Markets Hypothesis (Fama 1970), which states that it is impossible to outperform the market due to the informational efficiency of markets. 

While there are some similarities in market structure between financial markets and casino sports books, there are also several key differences (Levitt 2004). In financial markets, price is dictated by supply and demand. However, in sports betting markets, each casino dictates the \enquote{price} (odds ratio) at which the \enquote{stock} (specific bet) will trade. Levitt stresses the importance of the casino providing the right price for each bet by mentioning that the bookmaker can be exposed to significant losses if the price set by the bookmaker is incorrect. Another key difference between financial markets and sports betting markets is that the Securities and Exchange Commission requires brokers to execute trades at the National Best Bid and Offer, whereas it is up to the individual bettor to discover which casino is offering the best risk-reward ratio. 

As Moskowitz (2015) notes, several papers\footnote{\label{Many Papers} Snyder (1978), Vergin and Scriabin (1978), Ali (1979), Losey and Talbott (1980), Hausch et al (1981), Asch et al (1984), Zuber et al (1985), Thaler and Ziemba (1988), Sauer et al (1988), Gandar et al (1988), Camerer (1989), Golec and Tamarkin (1991), Woodland and Woodland (1994), Dare and MacDonald (1996), Gray and Gray (1997), Avery and Chevalier (1999), Kuypers (2000), and Lee and Smith (2002) have commented on the efficiency of various sports betting markets.} study the efficiency of sports betting markets, with the evidence somewhat mixed. We define an inefficient sports betting market as a market in which a betting strategy consistently generates profit. (Snyder 1978; Sauer et al 1988; Constantinou et al 2013). From our analysis, we find that there are exploitable inefficiencies due to differences in price offerings by individual casinos. Unlike any of the aforementioned papers (which tend to focus on one or two sports), we utilize a single pricing model to identify mispriced bets across several sports (namely the NFL, NBA, NCAAF, NCAAB, and WNBA). We extend Levitt's conclusion that bookmakers are more skilled at predicting game outcomes than bettors to the other major sports that we examine in this paper. This assumption that bookmakers are the most skilled at predicting game outcomes is key to formulating our model as we chiefly use information and data gathered from the entire market of casinos to take advantage of the mispricing by inefficient casinos. Due to our successful identifications of mispricings in sports betting markets, we were able to build a betting strategy that consistently generates profit across several different sports.

\section{Terminology} \label{sec:terminology}

Before diving deeper into the subject matter, we find it essential to define some of the parlance involved with sports betting. 

\subsection{Point Spread \& Spread Bets}

To paraphrase Stern (1991), the perceived difference between two sporting teams is measured by the point spread. For example, if Team A is determined to have a point spread of -6.5 when matched up against Team B, then Team A is favored to beat Team B by 6.5 points. In this scenario, Team A is the favorite (as they are favored to win) and Team B is the underdog. A bet placed on Team A will be a winning bet if and only if Team A beats Team B by more than 6.5 points. Conversely, Team B will have an inverse point spread of +6.5 points. A bet placed on Team B will be a winning bet if and only if Team B either wins \textit{or} loses by less than 6.5 points. The betting in this scenario is an example of a spread bet. A spread bet is designed to have even odds, meaning that chance of success and failure are equal at 50\%. For this reason, a spread bet generally pays out 91 cents for every dollar wagered if the bet is successful. The casino collects the other 9 cents, which is called a \enquote{vigorish}. If there is an equal amount of money on either side, then the casino collects 9 cents for each dollar risked, and is exposed to no risk.

Spreads can also be understood in the following manner, where $PFS$ represents the Projected Final Score of each team.

\begin{equation}
\text{Point Spread} = PFS_{Favorite} - PFS_{Underdog}
\end{equation}

Spreads often include the half point \enquote{hook} in order to avoid ties, as most major sports (basketball, football, soccer, hockey, etc.) do not have fractional point scores. In the event that the spread is a whole number (i.e. Team A is +6), and the outcome is Team A wins by 6 points, money is returned to anyone who bets on the game. This kind of event is called a \enquote{push}.

\subsection{Moneyline Bets}

The Moneyline bet is a wager on which team will win. Moskowitz mentions that Moneyline bets pay out various amounts of dollars depending on which team is bet. (Moskowitz 2015). For example, if the moneyline for Team A is -500 and the moneyline for Team B is +300, a successful wager of 500 dollars on Team A nets a 100 dollar profit while a successful wager of 100 dollars on Team B nets a 300 dollar profit. Our algorithm solely focuses on identifying mispriced moneyline bets.

\subsection{Market Return}

Defining average market return in the sports betting context is important since this activity does not lend itself to traditional financial return concepts, such as a comparison to given benchmarks. A bettor's average return in this market will be equal to the mean return of their wagers, weighted for bet sizes. Table \ref{random_results} reveals that a bettor who randomly wagers on spread bets will have a mean return of approximately -4.4\% for all sports, and a bettor who randomly wagers on moneyline bets will have a mean return between -6.4\% and 0.8\% for all sports. As 0\% is included in all of the randomized moneyline bet intervals for all sports, we view the average market return as equal to 0\%. With this definition of average market return, we define above average market return for sports betting to be long run positive returns. Table \ref{random_results} will be discussed in further detail in the Results section.

\section{Model \& Data} \label{sec:data}

\subsection{Probability and Expected Value} \label{sec:prob}

There are 16 casino sportsbooks in Las Vegas that we obtained data from. For any given sports game, at most 16 casinos will provide point spreads and moneyline betting odds. The point spread and moneyline odds vary from casino to casino. Thus, from our obtained data, we end up with a range of point spreads and moneyline odds. As there are are a plethora of teams in College Basketball and College Football, we only considered games where one team was a member of a \enquote{Power Five} conference\footnote{\label{Power Five Teams} The Power Five Conferences are defined to be the Atlantic Coast Conference, Big Ten Conference, Big 12 Conference, Pac-12 Conference, and Southeastern Conference.}. For all other sports, we considered all games between all teams, excluding pre-season (exhibition) games but including post-season (i.e. playoff) games.

Our algorithm utilizes this range of point spreads to determine the probability of a team winning the game. Unlike Stern, we do not assume that a team's implied probability of winning a game from the point spread is normally distributed. One limitation of Stern's model is its ability to only price games with the absolute value of the spread less than six. Our model is non-parametric and allows us to price moneyline bets for any point spread. The following equation describes our Bayesian formula for a team's likelihood to win a game, at point spread PS.

\begin{equation}
\mathbf{P(\textbf{Win} \ \vert \  PS) = \textbf{Historical Win} \ \textbf{\%}  \ \textbf{of Teams with Point Spread = }  \ \textbf{PS}}
\end{equation}

This formula is the engine of our algorithm. We theorize that a team at point spread PS will win at the same rate at which historical teams at spread PS have won. Results that are considered in determining the rate at which historical teams win only include games that have concluded before the game that we wish to price. If we know the probability of an event, and we know the payout for making a correct wager, we can define expected value for a \underline{one dollar bet} as follows.

\begin{equation}
\textbf{Expected Value} = [\textbf{P(Winning)}*\textbf{Payout}] - \textbf{P(Losing)}
\end{equation}

The expected value of a one dollar bet is defined as the probability of winning multiplied by the payout for winning, subtracted by the probability of losing multiplied by the money wagered. Since the money wagered in a one dollar bet is 1, the expected winnings are now solely subtracted by the probability of losing. This is how Equation 3 is formed. The expected value of a bet symbolizes the expected money made or lost from the bet. Theoretically, selecting bets that have a positive expected value, and betting on them, would generate profit over time. This is the foundation of our betting algorithm.

Clearly, calculating the probability and expected value requires the spreads and payouts from Vegas casinos and historical spread data and results (i.e. which teams won and lost). Vegasinsider.com has spread and payout data from Vegas casinos for many sporting events, as well as the results of these events. In order to properly price a game, we need the historical win percentage for teams at various spreads. From a variety of sites, we obtained historical spread data and results ranging from 1990 to 2017.

\subsection{Algorithms}

As mentioned earlier, a range of point spreads and moneylines are obtained from vegasinsider.com. As we assume that the bookmakers know how to predict game outcomes best and are incentivized to provide accurate prices (resulting in the market as a whole being informationally efficient), we prefer to obtain the most updated spreads and moneylines. Our data analysis agreed with this assumption, and revealed that bookmakers' last update (i.e. whether a player was injured, etc.) generally occurred within an hour of the game. From this finding, we decided that the casino's last update was the most accurate indicator of the various casino specified variables (point spread and moneyline payout), as it was closest to game time and thus contained the most recent information.

We have two slightly different strategies that utilize the range of spreads to find the probability of winning, which we will refer to as \enquote{Simple} ({Algorithm \ref{simp_algo}}) and \enquote{Weighted} ({Algorithm \ref{weighted_algo}}). Both algorithms utilize an input mapping of spreads to their respective probabilities, but the Weighted algorithm also uses an input mapping of spreads to their respective frequencies. The simple algorithm returns the probability of a team winning via calculating the simple average of the probabilities of the unique spreads. The weighted algorithm returns the probability of a team winning via calculating the weighted average of the probabilities of the spreads by using the frequency of the spread as the weighting factor. Below, {Algorithm \ref{simp_algo}} and {Algorithm \ref{weighted_algo}} show the pseudocode of each algorithm.

\[
\textbf{Place Algorithm \ref{simp_algo} Here} 
\]
\[
\textbf{Place Algorithm \ref{weighted_algo} Here} 
\]

With the probability determined using either strategy, and the payout determined as the maximum of the moneyline payouts, we can utilize the expected value formula in Equation 3 to find the expected value of picking a team to win a game. If our probability calculations are correct, and thus our expected value computations, we should make money by wagering on bets with positive expected values.

\subsection{Epsilon and Expected Value Thresholds}

As the returns in Table \ref{algo_results} will demonstrate, simply betting on all games with positive expected value generates positive return on investment for three out of five sports when the simple algorithm is used to calculate probability of winning, and five out of five sports when the weighted algorithm is used to calculate the probability of winning. This finding implies that there are inefficiencies in the market, as there exists a method of obtaining positive returns. For reference, randomized betting returns on the same data are provided in Table \ref{random_results}. Both of these tables will be analyzed in further detail in the Results section of the paper. From data analysis, we were able to make improvements on our model.

The first improvement on our model regards the issue of predicting games with large spreads. This scenario is why Stern's model works best pricing bets with the absolute value of the spread less than six, as the tails are hard to predict. If the spread is large, it implies that one team is heavily favored to beat the other team, and our model lacks the historical sample size at such large spreads to create a realistic probability estimate. Additionally, casinos make the payouts for betting on underdogs in this scenario ludicrously high, resulting in the model generating a high expected value for underdogs in these cases. 

To account for the effect of these games in the tails, we created the \enquote{Epsilon Threshold}. The goal of the Epsilon Threshold is to provide a symmetric set of bounds for the probabilities of teams winning games that we are willing to bet on when the bet has positive expected value. For bets outside the bounds, we always go with the probability favorite. After determining which Epsilon Value yields the highest total return on investment, we can define the optimal bettable range of winning percentages as: $(.5 - \epsilon, .5 + \epsilon)$. The optimal Epsilon value varies significantly from sport to sport. 

The second improvement we designed is the \enquote{Expected Value Threshold}. Expected Value Threshold is the the value of Expected Value that a bet must be greater than in order to place the bet. The default Expected Value Threshold that we use is 0. In this default case, any bet that has a positive Expected Value will be bet on. In general, as the Expected Value Threshold increases so does the return on investment, which leads us to believe that the model is detecting large inefficiencies very well.

\subsection{Betting Algorithm}

The Betting Algorithm (Algorithm {\ref{betting_algo}}) decides which team to bet on by examining the expected value of the bet and applying the epsilon and expected value thresholds. The algorithm takes the following inputs: the expected value of betting on the favorite, the expected value of betting on the underdog, the probability that the favorite wins, the probability that the underdog wins, the value of the epsilon hyper-parameter, and the value of the expected value threshold. The output of the algorithm is a decision to bet on the favorite, underdog, or neither team depending on the filtering criteria of expected value, epsilon threshold, and expected value threshold. Below, {Algorithm \ref{betting_algo}} shows the pseudocode of the algorithm.

\[
\textbf{Place Algorithm \ref{betting_algo} Here} 
\]

Return on Investment (ROI \%) is computed outside of the algorithm by comparing the algorithm's bets to the real world outcomes. If the algorithm picks a team that wins, the winnings from that bet are equivalent to the payout of the correct bet. Else, if the algorithm picks a team that loses, the losses from that bet are negative one (as every bet is a one dollar bet). Else, if the algorithm doesn't pick a team, there is no bet, so there are no winnings or losses. The ROI \% is the one hundred times sum of the winnings and losses divided by the number of bets.

\section{Results} \label{sec:result}

In order to provide benchmarks for our algorithm returns, we will first discuss Table \ref{random_results}, which displays the returns from randomly placing spread and moneyline bets. Next, we will contrast these results with the findings from Table \ref{algo_results}, which displays the returns for our two win probability models: Simple and Weighted. Finally, we will showcase a year by year, sport by sport breakdown of the optimal ROI with \ref{algo_results}

\[
\textbf{Place Table \ref{random_results} Here} 
\]

Table \ref{random_results} summarizes returns from three different strategies of randomly betting on games. Algorithms \ref{spread_random_betting_algo} and \ref{ml_random_betting_algo} describe the pseudocode that implements each of these strategies. The confidence intervals in each of the panels are 95th percentile confidence intervals, and each strategy involves betting on every game in the sample data. From Panel A, it's evident that randomly picking spread bets (regardless of sport) returns a mean ROI of around -4.4\%. In this randomization approach, there is an equal chance the strategy will choose either team to cover the spread. In Panel B, the mean ROIs have more variation than Panel A, and the mean returns are close to 0\% (excluding NCAAB). In the randomization approach used to generate Panel B, the favorite and the underdog have an equal chance to be picked to win the game. In Panel C, the mean ROIs are similar to Panel C's, but generally smaller (excluding NCAAB). The randomization approach used to generate Panel C assigns a 67\% chance that the favorite will be chosen to win the game, and a 33\% chance that the underdog will be chosen to win the game. 

From comparing all three panels, it's clear that randomly placing moneyline bets generally have a larger returns on investment than randomly placing spread bets. From these results, we chose to build our betting strategies around successfully pricing and placing moneyline bets. Additionally, it's notable that all of the 95\% confidence intervals (excluding NBA spread bets from Panel A) contain 0 in the interval, as we have defined an inefficient market as a market where a strategy can consistently generate profit. Clearly, we cannot reject the null hypothesis that sports betting markets are jointly efficient for all of these sports for any kind of bet. 

\[
\textbf{Place Table \ref{algo_results} Here} 
\]

This table summarizes returns from two win probability models: Simple and Weighted. Algorithms \ref{simp_algo} and \ref{weighted_algo} describe the pseudocode that implements each of these models. Algorithm \ref{betting_algo} describes how these probability models are incorporated into the betting process. Both Panel A and B follow the same structure. First, returns are reported for bets made sheerly on an expected value basis derived from the specific win probability model. Next, returns are reported for the betting algorithm with the Epsilon Threshold parameter. Finally, returns are displayed for the betting algorithm augmented with the Expected Value Threshold parameter as well as the Epsilon Threshold. 

Returns always increase at each step, as we have selected the optimal values of the Epsilon and Expected Value Thresholds. We found these optimal parameter values by maximizing Total Return, which is defined as ROI * N, where N is the number of bets. In Panel A, the returns resulting from strictly making positive expected value bets according to the Simple model vary across sports, with the NCAAF and NCAAB having negative ROI. In comparison, the returns from making positive expected value bets according to the Weighted model also vary across sports, but are always positive. As aforementioned, returns increase when more parameters are included. The best conclusion to draw from comparing the returns augmented with optimal Epsilon and EV Thresholds for both algorithms is that neither model really dominates the other. This leads us to believe that choosing Simple or Weighted on a sport by sport basis is probably the best idea. Next, we'll do a year by year breakdown of each sport's ROI to provide deeper detail and context for the results.

\[
\textbf{Place Table \ref{yearly_roi} Here} 
\]

Table \ref{yearly_roi} shows a year by year, sport by sport breakdown of optimal ROI (bets made by betting algorithm augmented with optimal Epsilon and Expected Value Threshold parameters), with the S\&P 500 ROI listed as a benchmark. The All Leagues column shows a weighted ROI generated by summing the Total Return (ROI * Number of Games) for each sport, and then dividing by the total number of games bet upon for each year. We do not benefit from compounding, as we designate each bet as an independent \$ 1 bet. Additionally, even though we are computing the necessary inputs, we do not utilize Kelly's criterion.

The sample of games we have excludes the early 2000's, 2008, and 2018, which are the years in the 21st century in which S\&P 500 had a negative return. In general, the S\&P 500 outperforms the sports betting returns (excluding 2012 for Simple, and 2011 and 2015 for both strategies). We present this table as the foundation of the argument that sports betting should be viewed as an alternative asset management strategy, as it logically has no relation to the stock market (market neutral). The next table will show the yearly sample size, and we will explain why we are unable to collect data for more past years.

\[
\textbf{Place Table \ref{yearly_roi_sample_size} Here} 
\]

This table supports Table \ref{yearly_roi}, and provides the sample size of bets for each sport in each year. In general, as time goes on, we bet on more games. This is a feature of the data as we had to web scrape most of the line movements of the games from an internet archive (archive.org/web) as the site the data were originally found at (vegasinsider.com) does not display line movement data for any games outside of the current year, and it is challenging to find older line movement data. We are also limited by our database of line movements. In certain sports (WNBA), sufficient historical spread win probability data is lacking due to the sparsity of both games played and line movements measured. An interesting result of this comparison of tables is that the sample size of the Simple win probability model tends to be considerably larger for more recent years of data.

Comparing the results from Table \ref{random_results} and Tables \ref{algo_results} and \ref{yearly_roi} highlights that the optimal ROI's from Table \ref{algo_results} are much larger than the mean ROI from Table \ref{random_results}. To show that the expected optimal ROI's are not a feature of over-fitting the hyper-parameters (Epsilon Value and Expected Value Threshold) to the sample data, we utilized the bootstrap procedure. If we can show that the expected optimal ROI's for the betting strategy are larger than zero with statistical significance, we have proved that the strategy consistently produces a profit. Proving that the strategy consistently produces a profit clearly indicates that the sports betting markets are inefficient.

\subsection{Bootstrap Results}

In order to demonstrate the robustness of our results, we utilized the bootstrap procedure. The algorithm that describes our implementation is Algorithm \ref{bs_algo}. This algorithm is utilized for each sport, and the number of times the data is resampled with replacement for each sport is equivalent to the number of sample games for that sport. 10,000 iterations of this algorithm are used to generate the desired bootstrap samples from which Optimal ROI, Optimal Epsilon Value, Optimal Expected Value Threshold, and Optimal Number of Games can be computed. Univariate and bivariate visualizations of the aformentioned variables in the bootstrap samples are available in the Online Supplement. The univariate visualizations are histograms with optimal bin sizing calculated by Scott's Rule (Scott 1979), and the bivariate visualizations are contour plots with densities calculated via averaged shifted histograms (Scott 1985). Both visualization techniques are non-parametric.

From the bootstrap samples, Table \ref{bs_95} contains 95\% Confidence Intervals for each variable sampled. From the univariate visualizations in the Online Supplement, it's evident that some of the samples are clearly skewed. In order to handle the skewness of the data, the confidence intervals are calculated by two different non-parametric procedures: Percentile Intervals (Efron 1993) and High Density Intervals (Hyndman 1996). 

\[
\textbf{Place Table \ref{bs_95} Here} 
\]

Table \ref{bs_95} displays two-sided 95\% confidence intervals for the bootstrapped data.  Panel A and B highlight the intervals for bets relying upon the Simple win probability model, and Panel C and D highlight the intervals for bets relying upon the Weighted win probability model. Both  Panel  A  and  C  utilize  95\%  percentile  intervals, and  Panel B and D utilize  95\%  high density intervals. All ROI and Epsilon Value intervals for all sports in each panel do not contain zero. 

To examine whether the optimal ROI values (or optimal Epsilon Values) for all sports are statistically significant and greater than zero, we must test the following set of joint hypotheses: $H_0 = 0$ and $H_A > 0$. In order to perform the joint hypothesis test (as we are checking these hypotheses for all sports concurrently), we use the Bonferroni correction with $\alpha = 0.05$. Since we are testing five sports concurrently with the aforementioned hypotheses for each win probability model, we must now examine whether all of the one sided 99\% confidence intervals contain zero for each win probability model. 

\[
\textbf{Place Table \ref{bs_99} Here} 
\]

Table \ref{bs_99} displays one-sided 99\% confidence intervals for bootstrapped data. Panel A highlights the intervals for bets relying upon the Simple win probability model, and Panel B highlights the intervals for bets relying upon the Weighted win probability model. All ROI intervals for all sports in each panel do not contain zero. As we are utilizing an one-sided interval, the null and alternative hypotheses for ROI, EV Threshold, and Epsilon Value are respectively: $H_0 = 0$ and $H_A > 0$. By examining the intervals in the table and using the Bonferroni correction (with $\alpha = 0.05$) for each win probability model, we reject the null hypothesis that the optimal returns from this strategy are equivalent to zero for every sport. By again examining the intervals in the table and using the Bonferroni correction (with $\alpha = 0.05$) for each win probability model, we reject the null hypothesis that the optimal Epsilon Value is equivalent to zero for every sport. From the above, we have shown that the optimal ROI is statistically significant and greater than zero, meaning that there is a betting strategy that consistently generates profit.

\section{Conclusion} \label{sec:conclusion}

Levitt (2004) succinctly states that there is little evidence of bettors who are able to beat the bookies systematically. For horseracing bets, Asch et al (1982) were not able to devise a successful strategy based solely on observable betting odds. Through our analysis, we have shown that we have created a betting algorithm that solely relies upon betting odds and is capable of generating profit for multiple sports concurrently. If the market is efficient, there should not be any skillful way to generate above average market returns in the long run. Through our research and successful algorithm, we demonstrated the existence of market inefficiencies in sports betting markets. Our paper is the first to propose a single theory which identifies inefficiencies across sports through generating positive returns. 

We encourage future researchers to further explore the benefits and costs of using compounding bets rather than independent \$1 bets that we used in this paper, as well as properly utilizing Kelly's portfolio growth criterion (Kelly 1956). In addition, we also leave it to future researchers to propose improvements on our win probability models and betting algorithm parameters.

As the Supreme Court of the United States has struck down a 1992 law prohibiting sports betting\footnote{Murphy v. National Collegiate Athletic Association, 584 U.S. \_\kern-.5ex\_\kern-.5ex\_\kern-.5ex\_\kern-.5ex\_ 2018}, we anticipate that betting markets in the United States will continue to grow larger and more popular. Indeed, several states\footnote{New Jersey, Delaware, Mississippi, and West Virginia are a few examples.} have already legalized sports betting since the monumental decision. For now, we urge bettors to take advantage of these newfound opportunities and enjoy beating the house. 

\newpage
\nocite{*}
\bibliographystyle{jfe}
\bibliography{beating_the_house}

\newpage

\begin{table}[!htbp]
\centering
\caption{\textbf{Randomized Betting Returns (\$1 Bets)}}
\resizebox{\linewidth}{!}{%
\begin{tabular}{lccccc}
\multicolumn{6}{l}{\textbf{Panel A: Equally Randomized Spread Betting Returns}}\\
  \toprule
\textbf{} & \textbf{NFL} & \textbf{NBA} & \textbf{NCAAF} & \textbf{NCAAB} & \textbf{WNBA}\\ 
\midrule
Total Games Bet & 922.00 & 4019.00 & 951.00 & 1140.00 & 255.00 \\ 
Mean Return on Investment (ROI) \% & -4.35 & -4.39 & -4.41 & -4.39 & -4.39 \\ 
95\% ROI Confidence Intervals & (-10.31, 1.49) & (-7.27, -1.43) & (-10.26, 1.39) & (-9.80, 1.08) & (-15.94, 7.27)\\
\bottomrule
\end{tabular}}
~\\~\\
\resizebox{\linewidth}{!}{%
\begin{tabular}{lccccc}
\multicolumn{6}{l}{\textbf{Panel B: Equally Randomized Moneyline Betting Returns}}\\
  \toprule
\textbf{} & \textbf{NFL} & \textbf{NBA} & \textbf{NCAAF} & \textbf{NCAAB} & \textbf{WNBA}\\ 
\midrule
Total Games Bet & 922.00 & 4019.00 & 951.00 & 1140.00 & 255.00 \\ 
Mean Return on Investment (ROI) \% & 0.75 & 0.54 & 0.30 & -6.44 & -0.18 \\ 
95\% ROI Confidence Intervals & (-6.48, 7.90) & (-3.26, 4.23) & (-8.14, 8.59) & (-12.97, 0.09) & (-14.62, 14.07) \\
\bottomrule
\end{tabular}}
~\\~\\
\resizebox{\linewidth}{!}{%
\begin{tabular}{lccccc}
\multicolumn{6}{l}{\textbf{Panel C: Unequally Randomized Moneyline Betting Returns}}\\
  \toprule
\textbf{} & \textbf{NFL} & \textbf{NBA} & \textbf{NCAAF} & \textbf{NCAAB} & \textbf{WNBA}\\ 
\midrule
Total Games Bet & 922.00 & 4019.00 & 951.00 & 1140.00 & 255.00 \\ 
Mean Return on Investment (ROI) \% & 0.72 & -1.04 & -1.52 & -3.78 & -1.39 \\ 
95\% ROI Confidence Intervals & (-6.22, 7.56) & (-4.47, 2.39) & (-8.49, 5.67) & (-9.48, 1.92) & (-14.35, 12.25) \\
\bottomrule
\end{tabular}}
\label{random_results}
\end{table}

\singlespacing{\noindent \footnotesize
This table summarizes returns from three different strategies of randomly betting on games. Algorithms 4 and 5 describe the pseudocode that implements each of these strategies. The confidence intervals in each of the panels are 95th percentile confidence intervals, and each strategy involves betting on every game in the sample data. From Panel A, it's evident that randomly picking spread bets (regardless of sport) returns a mean ROI of around -4.4\%. In this randomization approach, there is an equal chance the strategy will choose either team to cover the spread. This approach mirrors the fact that spread bets are designed to have an equal chance of the favorite or underdog covering the spread. In Panel B, the mean ROIs have more variation than Panel A, and the mean returns are close to 0\% (excluding NCAAB). In the randomization approach used to generate Panel B, the favorite and the underdog have an equal chance to be picked to win the game. In Panel C, the mean ROIs are similar to Panel B's, but generally smaller (excluding NCAAB). The randomization approach used to generate Panel C assigns a 67\% chance that the favorite will be chosen to win the game, and a 33\% chance that the underdog will be chosen to win the game. From comparing all three panels, it's clear that randomly placing moneyline bets generally have a larger returns on investment than randomly placing spread bets. From these results, we chose to build our betting strategies around successfully placing moneyline bets.}

\newpage

\begin{table}[!htbp]
\centering
\caption{\textbf{Betting Algorithm Returns (\$1 Bets)}}
\resizebox{\linewidth}{!}{%
\begin{tabular}{lrrrrr}
\multicolumn{6}{l}{\textbf{Panel A: Simple Algorithm Results}}\\
  \toprule
\textbf{} & \textbf{NFL} & \textbf{NBA} & \textbf{NCAAF} & \textbf{NCAAB} & \textbf{WNBA}\\ 
  \midrule
\textbf{Simple} & & & & &\\
\midrule
Total Games Analyzed & 922.00 & 4019.00 & 951.00 & 1140.00 & 255.00 \\ 
Games Bet (Simple) & 742.00 & 2386.00 & 667.00 & 709.00 & 226.00 \\ 
Return on Investment (ROI) \% & 10.81 & 3.42 & -2.49 & -1.10 & 4.38 \\
\midrule
\textbf{Simple with Optimal Epsilon Threshold} & & & & &\\
\midrule
Epsilon Value & 0.34 & 0.38 & 0.25 & 0.17 & 0.20 \\ 
Bet Games & 742.00 & 2386.00 & 667.00 & 709.00 & 226.00 \\ 
ROI \% & 11.55 & 4.02 & 2.97 & 1.39 & 7.42 \\
\midrule
\textbf{Simple with Optimal Epsilon and EV Thresholds} & & & & &\\
\midrule
EV Threshold & 0.013 & 0.010 & 0.083 & 0.007 & 0.084 \\ 
Games Bet & 567.00 & 1500.00 & 292.00 & 645.00 & 151.00 \\ 
ROI \% & 16.57 & 9.04 & 7.29 & 4.07 & 17.01 \\ 
\bottomrule
\end{tabular}}
~\\~\\
\centering
\resizebox{\linewidth}{!}{%
\begin{tabular}{lrrrrr}
\multicolumn{6}{l}{\textbf{Panel B: Weighted Algorithm Results}}\\
  \toprule
\textbf{} & \textbf{NFL} & \textbf{NBA} & \textbf{NCAAF} & \textbf{NCAAB} & \textbf{WNBA} \\ 
  \midrule
\textbf{Weighted} & & & & &\\
\midrule
Total Games Analyzed & 922.00 & 4019.00 & 951.00 & 1140.00 & 255.00 \\ 
Games Bet (Weighted) & 766.00 & 2400.00 & 664.00 & 685.00 & 235.00 \\ 
Return on Investment (ROI) \% & 5.73 & 0.66 & 1.60 & 0.94 & 4.55 \\
\midrule
\textbf{Weighted with Optimal Epsilon Threshold} & & & & &\\
\midrule
Epsilon Value & 0.12 & 0.35 & 0.25 & 0.37 & 0.16 \\ 
Bet Games & 766.00 & 2400.00 & 664.00 & 685.00 & 235.00 \\ 
ROI \% & 8.69 & 1.57 & 6.06 & 2.67 & 9.70 \\
\midrule
\textbf{Weighted with Optimal Epsilon and EV Thresholds} & & & & &\\
\midrule
EV Threshold & 0.022 & 0.040 & 0.010 & 0.030 & 0.023 \\ 
Games Bet & 662.00 & 619.00 & 599.00 & 284.00 & 216.00 \\ 
ROI \% & 9.26 & 11.18 & 8.04 & 13.02 & 12.98 \\ 
   \bottomrule
\end{tabular}}
\label{algo_results}
\end{table}
\singlespacing{\noindent \footnotesize
This table summarizes returns from two win probability models: Simple and Weighted. Algorithms 1 and 2 describe the pseudocode that implements each of these models. Algorithm 3 describes how these probability models are incorporated into the betting process. Both Panel A and B follow the same structure with returns reported for bets made sheerly on an expected value basis derived from the specific win probability model. Next, returns are reported for the betting algorithm with the Epsilon Threshold parameter. Finally, returns are displayed for the betting algorithm augmented with the Expected Value Threshold parameter as well as the Epsilon Threshold. Returns always increase at each step, as we have selected the optimal values of the Epsilon and Expected Value Thresholds. We chose these parameter values to optimize for maximizing Total Return, which is defined as ROI * N, where N is the number of bets. In Panel A, the returns resulting from strictly making positive expected value bets according to the Simple model vary across sports, with the NCAAF and NCAAB having negative ROI. In comparison, the returns from making positive expected value bets according to the Weighted model also vary across sports, but are always positive. As aforementioned, returns increase when more parameters are included. The best conclusion to draw from comparing the returns with both algorithms augmented with optimal Epsilon and EV Thresholds is that neither model really dominates the other, which leads us to believe that choosing Simple or Weighted on a sport by sport basis is probably the best idea.}

\newpage

\begin{table}[ht]

\caption{\textbf{Yearly Optimal Return on Investment [\%] (\$1 Bets)}}
\resizebox{\linewidth}{!}{%
\begin{tabular}{ccccccccc}
\multicolumn{9}{l}{\textbf{Panel A: Simple Return on Investment}}\\
\toprule
Year & NFL & NBA & NCAAF & NCAAB & WNBA & All Leagues & Sample Size & S\&P 500\\
\midrule
2017 & 34.01 & 4.20 & 0.00 &-0.30 & 7.58&3.38 & 839 & 21.69\\
2016 & -1.56 & 8.56 &-6.57 &-0.22 & -3.83&2.22 & 624 & 11.80\\
2015 & 43.71 & 17.69 &26.12 &18.24 &107.79 &29.09 & 365 & 1.34\\
2014 & 4.03 & 2.13 &64.00 &-13.87 & 0.00&8.22 &164 & 13.53\\
2013 & -15.16 & 11.40 & -8.82& 16.25& 100.00&10.64 & 189 & 32.21\\
2012 & 9.77  & 25.11 &7.22 &8.33 & 77.78& 17.51 & 128 & 15.84\\
2011 & 76.03 & 16.79 & 12.00& ---& ---& 29.19 & 272 & 2.06\\
2010 & -7.89 & 6.07 & 0.85& ---& ---& 2.84& 384 & 14.93\\
2009 & 1.10 & 4.37 & 11.27& ---& ---& 4.25 & 190 & 26.42\\
\midrule
\end{tabular}}
~\\~\\
\resizebox{\linewidth}{!}{%
\begin{tabular}{ccccccccc}
\multicolumn{9}{l}{\textbf{Panel B: Weighted ROI Return on Investment}}\\
\toprule
Year & NFL & NBA & NCAAF & NCAAB & WNBA & All Leagues & Sample Size & S\&P 500\\
\midrule
2017 & 12.46 & -3.68 &-25.00  &3.29 & 7.38&3.28 & 465 & 21.69\\
2016 & 5.95 & 7.57 &5.59 &-6.74 &15.07 &5.23&475 & 11.80\\
2015 & 14.67 & 22.88 &10.77 & 23.95 &47.64 &18.35& 344 & 1.34 \\
2014 & -12.59 & 7.05 &51.40 &-10.83 & 0.00&6.17 & 128 & 13.53\\
2013 & -13.31 & 24.82 &-43.14 &-2.06 & 100.00&-0.89 & 121 & 32.21\\
2012 & -12.67 & 36.52 &9.34 &8.33 & 5.56&10.00 &150 & 15.84\\
2011 & 46.96 & 23.09 &11.05 &--- & ---&26.23 &199 &2.06\\
2010 & 17.60 & 11.65 &6.91 &--- & ---&11.55 &318 & 14.93\\
2009 & -0.20 & -2.88 &9.19 &--- & ---&2.15 &170 & 26.42\\
\midrule

\end{tabular}}
\label{yearly_roi}
\end{table}

\singlespacing{\noindent \footnotesize This table shows a year by year, sport by sport breakdown of optimal ROI (bets made by betting algorithm augmented with optimal Epsilon and Expected Value Threshold parameters), with the S\&P 500 ROI listed as a benchmark. The All Leagues column shows a weighted ROI generated by summing the Total Return (ROI * Number of Games) for each sport, and then dividing by the total number of games bet upon for each year. We do not benefit from compounding, as we designate each bet as an independent \$ 1 bet.  The sample of games we have excludes the early 2000's and 2008 and 2018, which are the years in the 21st century in which S\&P 500 had a negative return. In general, the S\&P 500 outperforms the sports betting returns (excluding 2012 for Simple, and 2011 and 2015 for both strategies). We present this table as the foundation of the argument that sports betting should be viewed as an alternative asset management strategy, as it logically has no relation to the stock market (market neutral).}

\newpage

\begin{table}[!htbp]
\centering
\caption{\textbf{Yearly Sample Sizes (Games Bet On)}}
\begin{tabular}{ccccccc}
\multicolumn{7}{l}{\textbf{Panel A: Simple Algo Sample Size (Games Bet On)}}\\
\toprule
Year & NFL & NBA & NCAAF & NCAAB & WNBA & All Leagues \\
\midrule
2017 & 16 & 343 & 0 &354 & 126& 839\\
2016 & 145 & 266 &90 &114 & 9& 624\\
2015 & 107 & 124 &29 &93 &12 & 365\\
2014 & 56 & 81 &17 &10 & 0 &164\\
2013 & 14 & 85 & 15& 73& 2& 189\\
2012 & 33  & 61 &31 &1 & 2&  128\\
2011 & 60 & 174 & 38& 0& 0&  272\\
2010 & 72 & 267 & 45& 0& 0&  384\\
2009 & 64 & 99 & 27& 0& 0&  190 \\
\midrule
\end{tabular}
~\\~\\

~\\~\\
\centering
\begin{tabular}{ccccccc}
\multicolumn{7}{l}{\textbf{Panel B: Weighted Algo Sample Size (Games Bet On)}}\\
\toprule
Year & NFL & NBA & NCAAF & NCAAB & WNBA & All Leagues \\
\midrule
2017 & 21 & 124 & 2 & 141 & 177 & 465\\
2016 & 165 & 95 & 158 & 44 & 13 & 475\\
2015 & 141 & 53 & 84 & 43 & 23 & 344\\
2014 & 64 & 34 & 27 & 3 & 0 & 128\\
2013 & 17 & 43 & 25& 34 & 2 & 121\\
2012 & 36  & 33 & 79 & 1 & 1 & 150\\
2011 & 62 & 66 & 71 & 0 & 0 & 199\\
2010 & 80 & 131 & 107 & 0 & 0 & 318\\
2009 & 76 & 40 & 54 & 0 & 0 & 170\\
\midrule
\end{tabular}
\label{yearly_roi_sample_size}
\end{table}

\singlespacing{\noindent \footnotesize This table supports Table \ref{yearly_roi}, and provides the sample size of bets for each sport in each year. In general, as time goes on, we bet on more games. This is a feature of the data as we had to web scrape most of the line movements of the games from an internet archive (archive.org/web) as the site the data were originally found at (vegasinsider.com) does not display line movement data for any games outside of the current year, and it is challenging to find older line movement data. We are also limited by our database of line movements. In certain sports, sufficient historical spread win probability data is lacking due to the sparsity of both games played and line movements measured. An interesting result of this comparison of tables is that the sample size of the Simple win probability model tends to be considerably larger for more recent years of data.}

\newpage

\begin{table}[ht!]
\caption{\textbf{Two Sided 95\% Bootstrap Confidence Intervals}}
\centering
\resizebox{\linewidth}{!}{%
  \begin{tabular}{ccccc}
  \multicolumn{5}{l}{\textbf{Panel A: Simple High Density Intervals}}\\
  \toprule
    \textbf{Sport } &
      {\textbf{ROI \%}} &
      {\textbf{EV Threshold}} &
      {\textbf{Epsilon Value}} &
      {\textbf{\% of Games Bet}}\\
    \midrule
    {NFL} & (7.04, 35.81) & (0.00, 0.03) & (0.11, 0.36) & (0.38, 0.82)\\ 
     {NBA} & (3.74, 16.00) & (0.00, 0.02) & (0.19, 0.46) & (0.28, 0.60) \\
    {NCAAB} & (1.36, 16.08) & (0.01, 0.03) & (0.14, 0.37) & (0.29, 0.60)\\ 
    {NCAAF} & (1.10, 16.00) & (0.00, 0.08) & (0.01, 0.30) & (0.32, 0.73) \\ 
    {WNBA} & (8.72, 57.19) & (0.00, 0.10) & (0.13, 0.44) & (0.37, 0.89)\\ 
  \end{tabular}}
~\\~\\
\centering
\resizebox{\linewidth}{!}{%
  \begin{tabular}{ccccc}
  \multicolumn{5}{l}{\textbf{Panel B: Simple Percentile Intervals}}\\
  \toprule
    \textbf{Sport } &
      {\textbf{ROI \%}} &
      {\textbf{EV Threshold}} &
      {\textbf{Epsilon Value}} &
      {\textbf{\% of Games Bet}}\\
    \midrule
    {NFL} & (9.40, 40.60) & (0.00, 0.06) & (0.10, 0.36) & (0.24, 0.81)\\ 
     {NBA} & (4.18, 16.87) & (0.00, 0.03) & (0.17, 0.46) & (0.21, 0.56) \\
    {NCAAB} & (2.07, 16.94) & (0.01, 0.03) & (0.14, 0.37) & (0.29, 0.60)\\ 
    {NCAAF} & (1.69, 17.08) & (0.00, 0.08) & (0.01, 0.30) & (0.31, 0.72) \\ 
    {WNBA} & (8.73, 47.45) & (0.01, 0.15) & (0.13, 0.32) & (0.35, 0.91)\\ 
  \end{tabular}}
  ~\\~\\
\centering
\resizebox{\linewidth}{!}{%
  \begin{tabular}{ccccc}
  \multicolumn{5}{l}{\textbf{Panel C: Weighted High Density Intervals}}\\
  \toprule
    \textbf{Sport } &
      {\textbf{ROI \%}} &
      {\textbf{EV Threshold}} &
      {\textbf{Epsilon Value}} &
      {\textbf{\% of Games Bet}}\\
    \midrule
    {NFL} & (7.04, 35.81) & (0.00, 0.03) & (0.11, 0.36) & (0.38, 0.82)\\ 
     {NBA} & (3.74, 16.00) & (0.00, 0.02) & (0.19, 0.46) & (0.28, 0.47) \\
    {NCAAB} & (1.36, 16.08) & (0.01, 0.03) & (0.14, 0.37) & (0.29, 0.60)\\ 
    {NCAAF} & (1.10, 16.00) & (0.00, 0.08) & (0.01, 0.30) & (0.32, 0.73) \\ 
    {WNBA} & (8.72, 57.19) & (0.00, 0.10) & (0.13, 0.44) & (0.37, 0.89)\\ 
  \end{tabular}}
  ~\\~\\
\centering
\resizebox{\linewidth}{!}{%
  \begin{tabular}{ccccc}
  \multicolumn{5}{l}{\textbf{Panel D: Weighted Percentile Intervals}}\\
  \toprule
    \textbf{Sport } &
      {\textbf{ROI \%}} &
      {\textbf{EV Threshold}} &
      {\textbf{Epsilon Value}} &
      {\textbf{\% of Games Bet}}\\
    \midrule
    {NFL} & (5.92, 41.32) & (0.00, 0.05) & (0.08, 0.34) & (0.30, 0.85)\\ 
     {NBA} & (1.66, 25.58) & (0.00, 0.04) & (0.12, 0.49) & (0.11, 0.60) \\
    {NCAAB} & (2.51, 19.49) & (0.00, 0.03) & (0.14, 0.38) & (0.23, 0.58)\\ 
    {NCAAF} & (2.03, 19.68) & (0.00, 0.05) & (0.01, 0.41) & (0.36, 0.73) \\ 
    {WNBA} & (6.43, 43.06) & (0.01, 0.15) & (0.13, 0.32) & (0.38, 0.93)\\ 
  \end{tabular}}
  \label{bs_95}
\end{table}

\singlespacing{\noindent \footnotesize This table displays two-sided 95\% confidence intervals for the bootstrapped data. Panel A and B highlight the intervals for bets relying upon the Simple win probability model, and Panel C and D highlight the intervals for bets relying upon the Weighted win probability model. Both Panel A and C utilize 95\% percentile intervals, and Panel B and D utilize 95\% high density intervals. All ROI and Epsilon Value intervals for all sports in each panel do not contain zero. }

\newpage

\begin{table}[ht!]
\caption{\textbf{One Sided 99\% Bootstrap Confidence Intervals}}
\centering
\resizebox{\linewidth}{!}{%
  \begin{tabular}{ccccl}
  \multicolumn{5}{l}{\textbf{Panel A: Simple Percentile Intervals}}\\
  \toprule
    \textbf{Sport } &
      {\textbf{ROI \%}} &
      {\textbf{EV Threshold}} &
      {\textbf{Epsilon Value}} &
      {\textbf{\% of Games Bet}}\\
    \midrule
    {NFL} & (8.11, 82.85) & (0.00, 0.10) & (0.06, 0.47) & (0.20, 0.84)\\ 
     {NBA} & (3.41, 40.25) & (0.00, 0.04) & (0.13, 0.50) & (0.14, 0.62) \\
    {NCAAB} & (1.20, 33.48) & (0.00, 0.03) & (0.04, 0.47) & (0.25, 0.66)\\ 
    {NCAAF} & (0.92, 38.44) & (0.00, 0.08) & (0.01, 0.46) & (0.27, 0.75) \\ 
    {WNBA} & (7.69, 145.12) & (0.00, 0.20) & (0.13, 0.50) & (0.30, 0.94)\\ 
  \end{tabular}}
~\\~\\
\centering
\resizebox{\linewidth}{!}{%
  \begin{tabular}{ccccc}
  \multicolumn{5}{l}{\textbf{Panel B: Weighted Percentile Intervals}}\\
  \toprule
    \textbf{Sport } &
      {\textbf{ROI \%}} &
      {\textbf{EV Threshold}} &
      {\textbf{Epsilon Value}} &
      {\textbf{\% of Games Bet}}\\
    \midrule
    {NFL} & (6.47, 111.64) & (0.00, 0.11) & (0.08, 0.37) & (0.27, 0.88)\\ 
     {NBA} & (1.77, 48.90) & (0.00, 0.04) & (0.07, 0.50) & (0.11, 0.62) \\
    {NCAAB} & (2.63, 35.02) & (0.00, 0.03) & (0.00, 0.46) & (0.27, 0.77)\\ 
    {NCAAF} & (2.34, 66.26) & (0.00, 0.08) & (0.01, 0.48) & (0.19, 0.75) \\ 
    {WNBA} & (6.80, 99.31) & (0.00, 0.22) & (0.05, 0.50) & (0.33, 0.96)\\ 
  \end{tabular}}
  \label{bs_99}
\end{table}

\singlespacing{\noindent \footnotesize This table displays one-sided 99\% confidence intervals for bootstrapped data. Panel A highlights the intervals for bets relying upon the Simple win probability model, and Panel B highlights the intervals for bets relying upon the Weighted win probability model. All ROI intervals for all sports in each panel do not contain zero. As we are utilizing an one-sided interval, the null and alternative hypotheses for ROI, EV Threshold, and Epsilon Value are respectively: $H_0 = 0$ and $H_A > 0$. By examining the intervals in the table and using the Boneferroni correction (with $\alpha = 0.05$) for each win probability model, we reject the null hypothesis that the optimal returns from this strategy are equivalent to zero for every sport. By again examining the intervals in the table and using the Boneferroni correction (with $\alpha = 0.05$) for each win probability model, we reject the null hypothesis that the optimal Epsilon Value is equivalent to zero for every sport.}


\newpage

{\SetAlgoNoLine
\begin{algorithm}[H]
\KwIn{$\pi$, a mapping of spreads to their respective probabilities}
\KwOut{Simple average of the probabilities of distinct spreads}
\BlankLine\BlankLine
Let $\sigma$ be the set of all unique spreads.\\
\Return $\frac{1}{|\sigma|}\sum_{s\in\sigma}\pi_{s};$
\caption{Compute Probability: Simple Average}
\label{simp_algo}
\end{algorithm}
}

\vspace{.25in}

{\SetAlgoNoLine
\begin{algorithm}[H]
\KwIn{$\phi$, a mapping of spreads to their respective frequencies\newline$\pi$, a mapping of spreads to their respective probabilities}
\KwOut{Average of the probabilities of all spreads weighted according to the spread frequencies}
\BlankLine\BlankLine
Let $\sigma$ be the set of all unique spreads.\\
\Return $\frac{\sum_{s\in\sigma}\pi_{s}\phi_{s}}{\sum_{s\in\sigma}\phi_{s}};$
\caption{Compute Probability: Weighted Average}
\label{weighted_algo}
\end{algorithm}
}


\vspace{.25in}

\begin{algorithm}[H]
\KwIn{$\bar{\mu}_{f}$, the expected value of winnings if the favorite team is bet on\\ \ \ \ \ \ \ \ \ \ \ \ $\bar{\mu}_{u}$, the expected value of winnings if the underdog team is bet on\\ \ \ \ \ \ \ \ \ \ \ \ $\pi_{f}$, the probability that the favorite team wins the game\\ \ \ \ \ \ \ \ \ \ \ \ $\pi_{u}$, the probability that the underdog team wins the game\\ \ \ \ \ \ \ \ \ \ \ \ $\epsilon$, value of the Epsilon hyper-parameter\\ \ \ \ \ \ \ \ \ \ \ \ $\tau$, value of the EV threshold hyper-parameter}
\KwOut{-1, if the algorithm decides to bet on neither team\\ \ \ \ \ \ \ \ \ \ \ \ \ \ \ 0, if the algorithm decides to bet on the underdog team OR \\ \ \ \ \ \ \ \ \ \ \ \ \ +1, if the algorithm decides to bet on the favorite team}
\BlankLine\BlankLine
\If{$\neg (\bar{\mu}_{f} < 0\ \wedge\ \bar{\mu}_{u} < 0)$} {
\uIf{$\pi_{f} \geq 0.50 + \epsilon$} {
\lIf{$\pi_{f} \geq \pi_{u}$} {
\Return 1
} \lElse {
\Return 0
}
} \Else {
\uIf {$\bar{\mu}_{f} \geq \bar{\mu}_{u}$} {
\lIf{$\bar{\mu}_{f} > \tau$} {
\Return 1
}
}
\Else {
\lIf{$\bar{\mu}_{u} > \tau$} {
\Return 0
}
}
}
}
$\Return\ -1;$
\caption{BettingAlgorithm}
\label{betting_algo}
\end{algorithm}


\newpage

\begin{algorithm}[htp]
\KwIn{
$\Gamma$, set of all games \\ \ \ \ \ \ \ \ \ \ \ \
$P_F$, points scored by favorite in each game\\ \ \ \ \ \ \ \ \ \ \ \
$P_U$, points scored by underdog in each game\\ \ \ \ \ \ \ \ \ \ \ \
$MS_F$, minimum spread for favorite in each game \\ \ \ \ \ \ \ \ \ \ \ \
$MS_U$, minimum spread for underdog in each game
}
\KwOut{Return on Investment (ROI \%)}
\BlankLine\BlankLine
\ForEach{$g \in \Gamma$}{    
    
    \uIf{$P_F + MS_F > P_U$}{
    $Victor \gets 1;$\
  }
  \uElseIf{$P_U + MS_U > P_F$}{
    $Victor \gets 0;$\
  }
  \Else{
    $Victor \gets 2;$\
  }
$Sim_{Victor} \sim U{[0,1]};$ \\
\uIf{$Sim_{Victor} > 0.5$}{
    $Sim_{Victor} \gets 1;$\
  }
  \Else{
    $Sim_{Victor} \gets 0;$\
  }
\uIf{$Victor = Sim_{Victor}$}{
    $Winnings \gets 100/110;$\
  }
  \uElseIf{$Victor \ \neq Sim_{Victor} \ \text{$\wedge$} \ Victor \ \neq 2$}{
    $Winnings \gets -1;$\
  }
  \Else{
    $Winnings \gets 0;$\
  }
}
ROI = $\frac{100}{|\Gamma|}\sum\limits_{g \in \Gamma}Winnings_{g};$\\
\Return{ROI}
\caption{SpreadBetRandomization}
\label{spread_random_betting_algo}
\end{algorithm}


\newpage

\begin{algorithm}[htp]
\KwIn{
$\Gamma$, set of all games \\ \ \ \ \ \ \ \ \ \ \ \
$P_F$, points scored by favorite in each game\\ \ \ \ \ \ \ \ \ \ \ \
$P_U$, points scored by underdog in each game\\ \ \ \ \ \ \ \ \ \ \ \
$PO_F$, payout for favorite in each game \\ \ \ \ \ \ \ \ \ \ \ \
$PO_U$, payout spread for underdog in each game \\ \ \ \ \ \ \ \ \ \ \ \
$\Theta$, probability of choosing favorite in each simulation
}
\KwOut{Return on Investment (ROI \%)}
\BlankLine\BlankLine
\ForEach{$g \in \Gamma$}{    
    
    \uIf{$P_F > P_U$}{
    $Victor \gets 1;$\
  }
  \uElseIf{$P_U > P_F$}{
    $Victor \gets 0;$\
  }
  \Else{
    $Victor \gets 2;$\
  }
$Sim_{Victor} \sim U{[0,1]};$ \\
\uIf{$Sim_{Victor} > \Theta$}{
    $Sim_{Victor} \gets 1;$\
  }
  \Else{
    $Sim_{Victor} \gets 0;$\
  }
\uIf{$Victor = Sim_{Victor} \ \text{$\wedge$} \ Victor = 1$}{
    $Winnings \gets PO_F;$\
  }
  \uElseIf{$Victor = Sim_{Victor} \ \text{$\wedge$} \ Victor = 0$}{
    $Winnings \gets PO_U;$\
  }
  \uElseIf{$Victor \ \neq Sim_{Victor} \ \text{$\wedge$} \ Victor \ \neq 2$}{
    $Winnings \gets -1;$\
  }
  \Else{
    $Winnings \gets 0;$\
  }
}
ROI = $\frac{100}{|\Gamma|}\sum\limits_{g \in \Gamma}Winnings_{g};$\\
\Return{ROI}
\caption{MoneylineBetRandomization}
\label{ml_random_betting_algo}
\end{algorithm}

\newpage

\begin{algorithm}[htp]
\small{
\KwIn{
$\Gamma$, set of all games \\ \ \ \ \ \ \ \ \ \ \ \
$N_G$, number of sample games \\ \ \ \ \ \ \ \ \ \ \ \
$EP$, vector of (0, $95^{th}$ Quantile) of Epsilon Values from bootstrap sample, step size of 0.01\\ \ \ \ \ \ \ \ \ \ \ \
$EV$, vector of (0, $95^{th}$ Quantile) of EV Thresholds from bootstrap sample, step size of 0.001 \\ \ \ \ \ \ \ \ \ \ \ \ 
}
\KwOut{
$O_{ROI}$, Optimal Return on Investment (ROI $\%$), \\ \ \ \ \ \ \ \ \ \ \ \ \ \ \
$O_{EP}$, Optimal Epsilon Value, \\ \ \ \ \ \ \ \ \ \ \ \ \ \ \
$O_{EV}$, Optimal Expected Value Threshold}
\BlankLine\BlankLine

$N_{BS} \gets 0;$\\
$\Gamma^{BS} \gets [\ ];$\\
\While {$N_{BS}\leq N_G$} {
randomly select a game, $g$ from $\Gamma;$ \\
$\Gamma^{BS}.append(g);$ \\
$N_{BS} \gets N_{BS} + 1;$}

initialize $TRBS$ as a matrix of zeros with dimensions ($|EP|$, $|EV|$); \\
initialize $ROIBS$ as a matrix of zeros with dimensions ($|EP|$, $|EV|$);\\
$i, j \gets 0, 0;$\\
\While{$i \leq |EP|$}{
	\While {$j \leq |EV|$} {
    	$N \gets 0;$\\
    	$TR \gets 0;$\\
       	\ForEach{$g \in \Gamma^{BS}$}{
          $D \gets BettingAlgorithm(g.\bar{\mu}_{f},\ g.\bar{\mu}_{u},\ g.\pi_{f},\ g.\pi_{u},\ EP_i,\ EV_j);$\\
          \lIf {$D \neq 0$} {$N \gets N + 1$}
          \If {$D = -1$} {
            \lIf {g.victor = Favorite} {$TR \gets TR - 1$}
            \lElse {$TR \gets TR + g.underdog\_payout$}
          }
          \ElseIf {$D = 1$} {
          	\lIf {g.victor = Favorite} {$TR \gets TR + g.favorite\_payout$}
            \lElse {$TR \gets TR - 1$}
          }
        }
        $TRBS[i, j] \gets TR$; \\
        $ROIBS[i, j] \gets \frac{TR}{N}$; \\
        $j \gets j + 1$ \\
	}
    $i \gets i + 1$\\
}
$i', j' \gets \underset{0\ \leq\ i\ \leq\ |EP| - 1,\ 0\ \leq\ j\ \leq\ |EV| - 1}{\arg\max}\ TRBS[i, j]$;\\
$O_{ROI} \gets ROIBS[i', j']$;\\
$O_{EP},\ O_{EV} \gets EP[i'],\ EV[j']$;\\

\Return{$O_{ROI}, O_{EP}, O_{EV}$};
}
\caption{BootstrapAlgorithm}
\label{bs_algo}
\end{algorithm}

\newpage

\end{document}